\documentclass{article}
\input epsf
\begin{document}
\begin{center}
SOFT PIONS AT HIGH ENERGY AND ITS PHENOMENOLOGICAL IMPLICATIONS\vspace{3mm}\\
S.KORETUNE\vspace{2mm}\\
Department of Physics, Shimane Medical University,
Izumo, Shimane 693-8501, JAPAN\\
E-mail: koretune@shimane-med.ac.jp
\end{center}
The soft pion theorem in the inclusive reaction at high energy
applied to the current induced reaction is explained briefly. 
A characteristic aspect of this theorem is the charge 
asymmetry produced by the pole terms in the soft pion limit. 
The pion charge asymmetry in the central region in the target-virtual-photon 
center of the mass (CM) frame of the semi-inclusive electroproduction and 
the contribution to the Gottfried sum are illustrated as examples.
\section{Introduction}
The soft pion theorem in the inclusive reaction at high energy was formulated by
Sakai and Yamada\cite{sa} many years ago. Compared with the soft pion theorem
in the exclusive reactions, the soft pion limit in the inclusive reaction
can not be directly related to the physical processes without an additional assumption
except in the neutral pion case. At present the fixed-mass sum rule approach\cite{kore78}
and the perturbative approach\cite{kore82} are known. The former approach
includes a new theoretical ingredient called as the current anticommutation relation
on the null plane\cite{kore80} and has been developed to the modified Gottfried sum 
rule\cite{kore93} and its relatives.\cite{kore98} The latter approach is an attempt 
to relate the structure function in the semi-inclusive reaction to the total 
inclusive reaction. By studying the semi-inclusive electroproduction,
it was suggested that we can identify the soft pion at high energy as the
directly produced pion with the low transverse momentum in the central region 
in the CM frame.\cite{kore78reso} Further, it was shown that the theoretically
expected value of the charge asymmetry in this kinematical region 
is very near to the experimental value.\cite{kore82}  
Thus this phenomena may have some relevance to the Gottfried 
sum.\cite{kore99} In this talk we explain these facts. 
\section{The current anticommutation relation on the null plane}
The connected matrix element of the current anticommutation relation 
on the null plane in the flavor $SU(3)\times SU(3)$ model takes the
following form. 
\begin{eqnarray}
\lefteqn{<p|\{J_a^+(x),J_b^+(0)\}|p>_c|_{x^+=0} } \nonumber \\
\lefteqn{= <p|\{J_a^{5+}(x),J_b^{5+}(0)\}|p>_c|_{x^+=0} } \nonumber \\
&=&\frac{1}{\pi }P(\frac{1}{x^-})\delta^2(\vec{x}^{\bot })
[d_{abc}A_c(p\cdot x , x^2=0) + f_{abc}S_c(p\cdot x , x^2=0)]p^+ . 
\end{eqnarray}
Intuitively we can see that a physical origin of the factor $P(1/x^-)$ lies in
the quantity $\partial \Delta^{(1)}(x)/\partial x^-$ at $x^+=0$. 
We can convince this fact very generally by using the Deser-Gilbert-Sudarshan(DGS)
representation\cite{DGS} for the current anticommutation relation 
which incorporates both the causality and the spectral condition.\cite{kore80,kore84}
Then, based on this equation, the modified Gottfried sum rule is derived as\cite{kore93}
\begin{eqnarray}
\lefteqn{\int^1_0\frac{dx}{x}\{F_2^{ep}(x,Q^2)-F_2^{en}(x,Q^2)\}}
\nonumber \\
 &=&\frac{1}{3}\left( 1-\frac{4f_{K}^2}{\pi}\int_{m_Km_N}^{\infty}\frac{d\nu}{\nu^2}
\sqrt{\nu^2-(m_Km_N)^2}\{\sigma^{K^+n}(\nu)-\sigma^{K^+p}(\nu)\}\right) .
\end{eqnarray}
This sum rule explains the NMC deficit in the
Gottfried sum\cite{NMC} almost model independently. It has shown that
the deficit is the reflection of the hadronic vacuum originating from the
spontaneous chiral symmetry breaking. In this sense the physics underlining
this algebraic approach has a common feature with that of the mesonic 
models reviewed in Ref.(12). However, in the algebraic approach,
importance of the high energy region not only in the theoretical
meaning but also in the numerical analysis has been made clear.
In fact, it shows that about $40 \%$ of the NMC deficit 
comes from the region where the momentum
of the kaon in the laboratory frame is above $4\;$GeV.
Further the numerical prediction based on this
sum rule exactly agrees with the recent experimental value from E866/NuSea
collaboration.\cite{E866} On the other hand, a typical
calculation in the mesonic model based on the $\pi NN$ and the $\pi N\Delta$ 
processes accounts for about a half of the NMC deficit,\cite{meso} and also
this model fairly well explains the experiment of E866/NuSea collaboration.
These facts suggest that there may exist a dynamical mechanism 
to produce the flavor asymmetry at medium and high energy which we have 
overlooked as yet, and that it may compensate the above flaw of the mesonic models.
\section{Soft pions at high energy}
Usually, the soft pion theorem has been considered to be applicable only in the low
energy regions. However in Ref.(1), it has been found that this theorem
can be used in the inclusive reactions at high energy if the Feynman's
scaling hypothesis holds. In the inclusive reaction ``$\pi + p \to
\pi_{s}(k) + anything$'' with the $\pi_{s}$ being the soft pion, 
it states that the differential cross-section
in the CM frame defined as
\begin{equation}
f(k^3,\vec{k}^{\bot},p^0)=k^0\frac{d\sigma}{d^3k} ,
\end{equation}
where $p^0$ is the CM frame energy, scales as
\begin{equation}
f\sim f^{F}(\frac{k^3}{p^0},\vec{k}^{\bot}) + \frac{g(k^3,\vec{k}^{\bot})}{p^0} .
\end{equation} 
If $g(k^3,\vec{k}^{\bot})$ is not singular at $k^3=0$, we obtain
\begin{equation}
\lim_{p^0\to \infty}f^F(\frac{k^3}{p^0},\vec{k}^{\bot}=0)=
f^F(0,0)=\lim_{p^0\to \infty}f(0,0,p^0) .
\end{equation}
This means that the $\pi$ mesons with the momenta $k^3<O(p^0)$ and
$\vec{k}^{\bot}=0$ in the CM frame can be interpreted as the soft pion.
This fact holds even when the scaling violation effect exists,
since we can replace the exact scaling by the approximate one in this
discussion.  The important point of this soft pion 
theorem is that the soft-pion limit can not be interchanged with the
manipulation to obtain the 
discontinuity of the reaction ``$a + b + \bar{\pi_s} \to a + b + \bar{\pi_s}$''. 
We must first take the soft pion limit in the reaction ``$a + b \to \pi_s + anything$''.
This is because the soft pion attached to the nucleon(anti-nucleon)
in the final state is missed in the discontinuity of the soft pion limit of
the reaction ``$a + b + \bar{\pi_s} \to a + b + \bar{\pi_s}$''. \cite{sa}
\section{The charge asymmetry}
\begin{figure}[h]
%\figurebox{20pc}{15pc}{} % to have a box alone
\epsfxsize=25pc % will enlarge or reduce the postscript figures based on the xsize
\epsfbox{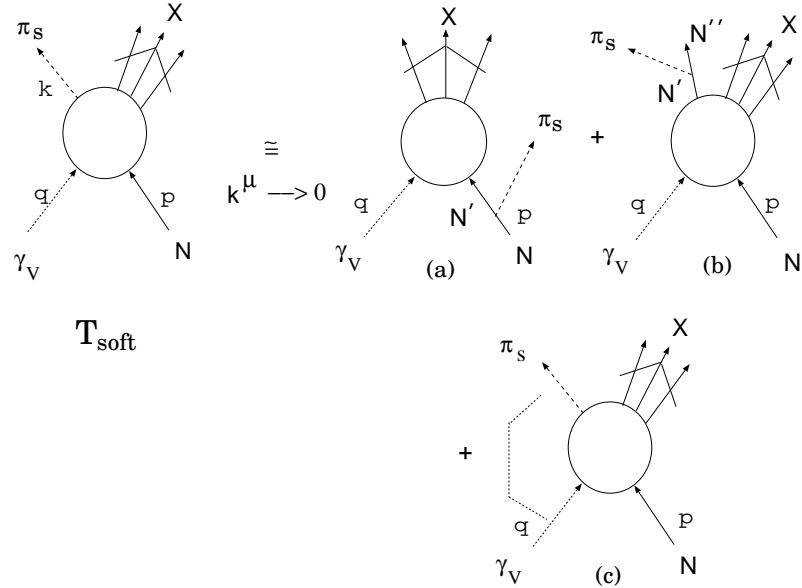} % postscript image file name
\caption{The amplitude in the soft pion limit. The graph (a) is the one coming from the soft pion
emission from the initial nucleon, the graph (b) is the one from the final nucleon(anti-nucleon),
and the graph (c) is the one from the commutation relation on the null plane.}
\end{figure}
The hadronic tensor is obtained by squaring the amplitude shown in Fig.1, and hence
we have many cross terms. Further in some cases the pole term
is inhibited by the charge conservation. Because of this fact, the application 
of the soft pion theorem in the inclusive reaction to
the charged pion case is non-trivial. In many cases the limit can not
be related to the known process without an additional assumption. 
Here we use the light cone current algebra\cite{fg} at some $Q^2_0$ where the evolution is started. 
We must take care whether the quantity is singlet piece or non-singlet one and
whether the quantity is charge conjugation even or odd.
Then the perturbative information can be taken into account through 
the structure functions related by this way. 
The usefulness of this method is that we can use the symmetry relations at $Q^2=Q_0^2$. 
Now the method in Ref.(1) had not been checked experimentally, it was
done in the soft $\pi^{-}$ case.\cite{kore78reso} From the experimental data
of the Harvard-Cornell group\cite{harvard} the data satisfying the following condition
are selected. 
\begin{enumerate}
\item[(1)]The transverse momentum satisfies $|\vec{k}^{\bot}|^2\leq m_{\pi}^2$.
\item[(2)]The change of $F^-$ can be regarded to be small in the small $x_F$ region.
\end{enumerate}
The effective cut of $x_F$ following the condition (2) is taken about at $0.2$. 
Then the theoretical value is roughly estimated as 10\% $\sim$ 20\% of the experimental 
value. However, in the central region, there are many pions from the decay of the 
resonances, and about 20\% $\sim$ 30\% can be expected to be the pion from the directly 
produced pion. Hence the theoretical value is the same order with the experimental value.
Now to reduce the ambiguity due to the pion from the resonance decay product,
the charge asymmetry was calculated.\cite{kore82} 
The theoretical value is roughly equal to $0.15 \sim 0.18$ with weak
$x_B$ dependence. While the experimental value\cite{harvard} with the transverse 
momentum satisfying the condition (1) is almost constant 
in the region $ 0< x_F < 0.1$ with its value $0.28 \pm 0.05$,
and it gradually decreases above $x_F = 0.1$. The data depends on $x_B$ weakly.
Hence the theoretical value is very near to the experimental value.
\section{The soft pion contribution to the Gottfried sum and its phenomenological implications}
The soft pion contribution to the Gottfried sum has been estimated.\cite{kore99}
Adding the contributions from the soft $\pi_s^+,
\pi_s^-,$ and $\pi_s^0$, subtracting the contributions to
$F_2^{en}$ from those to $F_2^{ep}$, and using symmetric sea polarization for simplicity,
we obtain
\begin{eqnarray}
\lefteqn{(F_2^{ep} - F_2^{en})|_{soft}}\nonumber \\
 &=& \frac{I_{\pi}}{4f_{\pi}^2}[g_A^2(0)(F_2^{ep} - F_2^{en})(3<n> -1)
-16xg_A(0)(g_1^{ep} - g_1^{en})] ,
\end{eqnarray}
where $I_{\pi}$ is the phase space factor for the soft pion and $<n>$
is the sum of the nucleon and anti-nucleon multiplicity.
To estimate the magnitude of this asymmetry, we approximate $F_2^{ep},
F_2^{en},g_1^{ep},g_1^{en}$ on the right-hand side of Eq.(6) by the
valence quark distribution functions at $Q_0^2=4\;$ GeV$^2$.\cite{GS}
As a multiplicity of the nucleon and antinucleon, we set
$<n> = a\log_es +1 $, where $s=(p+q)^2$. The parameter $a$ is fixed as
0.2 in consideration for the proton and the anti-proton multiplicity
in the $e^+e^-$ annihilation such that $\frac{1}{2}a\log_es$ with 
$\sqrt{s}$ replaced by CM energy of that reaction agrees 
with it.\cite{DELPHI}
Following the experimental check of the charge asymmetry in the previous section,
the transverse momentum is restricted by the condition (1) and the Feynman scaling
variable is cut at $x_F=0.1$. By calculating the phase space factor under these
conditions and allowing a small change of the parameters to determine
the phase space, it is shown that we can expect the magnitude of the contribution to 
the Gottfried sum from the soft pion is about $-0.04 \sim -0.02$. The main
contribution of the soft pion to this sum comes from the small $x_B$ region.\\
In the phenomenological parton model, we have not yet taken into account the soft
pion contribution except when we have a general constraint
such as the Adler sum rule. In such a case the soft pion contribution is
effectively taken into account in the quark distributions by 
satisfying the sum rule. However in case of the sea quark distributions,
no such sum rule constraint is imposed. We have many sum rules concerning
the sea quarks\cite{kore98,kore97} which have a clear physical meaning, and
the modified Gottfried sum rule is one example among these sum rules.
It gives us information on up and down sea quarks. The mean charge
sum rule for the light sea quarks stands on the same theoretical
footing as the modified Gottfried sum rule. Unfortunately,
all the presently available strange sea quark distribution badly breaks the 
mean charge sum rule,\cite{kore97} and here is a soft pion contribution which
has not yet been taken into account. In conclusion, combined analysis of the
sum rule and the soft pion at high energy gives us an important insight into the sea quark
distribution functions or the vacuum structure of the hadron.

\end{document}